\documentclass[12pt]{article}
\usepackage{amsfonts}
\usepackage{bbm}

\newcommand{\be}{\begin{equation}}
\newcommand{\ee}{\end{equation}}
\newcommand{\bea}{\begin{eqnarray}}
\newcommand{\eea}{\end{eqnarray}}
\newcommand{\nn}{\nonumber}
\newcommand{\lb}{\left[}
\newcommand{\rb}{\right]}
\newcommand{\ac}{\mathcal{A}}
\newcommand{\af}{\mathfrak{A}}
\newcommand{\bac}{\bar\mathcal{A}}
\newcommand{\ab}{\mathbb{A}}
\newcommand{\bab}{\bar\mathbb{A}}
\newcommand{\qc}{\mathcal{H}}
\newcommand{\bqc}{\bar\mathcal{H}}
\newcommand{\dc}{\mathcal{D}}
\newcommand{\bdc}{\bar\mathcal{D}}
\newcommand{\fc}{\mathcal{F}}
\newcommand{\fd}{\mathcal{L}}
\newcommand{\of}{{1\over 4}}
\newcommand{\hf}{{1\over 2}}
\newcommand{\ghf}{{g\over 2}}
\newcommand{\p}{\partial}
\newcommand{\vf}{\varphi}
\newcommand{\iu}{I}
\newcommand{\ou}{\mathit{1}}

\begin{document}
\begin{titlepage}

\title{Fermions from the gauge models ground state}

\author{M.N. Stoilov\\
{\small\it Bulgarian Academy of Sciences,}\\
{\small\it Institute of Nuclear Research and Nuclear Energy,}\\
{\small\it Blvd. Tzarigradsko Chausse\'e 72, Sofia 1784, Bulgaria}\\
{\small e-mail: mstoilov@inrne.bas.bg}}

\maketitle

\begin{abstract}
We investigate the quantization of pure $U(1)$ and $U(2)$ gauge theories in the vicinity of non-trivial  ground state in four-dimensional Euclidean space-time.
The main goal is to make the simultaneous consideration of many vacuums possible.
It is shown that Fueter (quaternion) analytic  and anti analytic functions 
can be used as vacuum's collective coordinates.
As a result the ground state describes not a single quasi  particle, or finite number of such particles, but a {\it field}.
This field satisfies the massless Dirac equation.
This is not a contradiction because it is known that massless spinors can be quantized either as fermions or as bosons.
We choose to quantize the vacuum anomalously (Fermi--Dirac).
The anomalous quantization of the  gauge fields ground state allows non-trivial (anti) self-dual configurations to exist.
The possible connection to the lepton sector of the Standard Model is discussed.
\end{abstract}

{\it
\vskip 10pt

PACS (2010): 11.15.Kc, 14.80.Hv
}

\end{titlepage}
\section{Introduction}

Quantization of a model  in the vicinity of a  solution of its equations of motion  is a common technique in Quantum Field Theory.
The field configuration around which we quantize
 represents the vacuum and the deviation of the 
field from this vacuum is the quantum field.
The consideration is very straightforward when the vacuum is unique.
But it is possible that the ground state is not unique, e.g., it may
depend on some parameters.
In this case, first,  a new gauge symmetry emerges in the model \cite{t1, t2, t3} and second, an alternation of  the standard path integral over all fields is required --- the functional integral is now   over the quantum field (with  measure which takes into account the new gauge symmetry)
plus additional ordinary integrals over all parameters (collective coordinates)
on which the vacuum depends.
Starting with the pioneer works \cite{gt},\cite{cb},\cite{vz},
the collective coordinate are widely used in the estimation of 
some  non-perturbative effects in different gauge models.

Here we follow an approach which diverts from the canonical collective coordinates prescription.
The reason is that we do not know the explicit functional form of the vacuum for the model we consider.
Instead of this we know that the ground state satisfies some well 
defined condition which does not coincide  with the equation of motion. 
The approach is suitable for ground state which is so degenerate that it depends on  
infinite many parameters, i.e. it is a field --- the vacuum field.
The question is how the functional integral looks like
in this case?
The answer, we shall argue for, is that the functional integral is on both quantum and vacuum fields.
In other words we quantize the vacuum as well.

Our basic example is the pure $U(1)$ gauge model in  four-dimensional Euclidean 
space-time $E^4$ and the classical solution around which we decompose the 
gauge potential is (anti) self-dual.
This seems rather trivial because it is known that for simple topological reasons there are no instanton configurations in Electrodynamics. 
However, we find a way out to bypass the problem.
Let us forget for a moment what we know about $U(1)$ instantons and consider the conditions for (anti) self-duality.
Using quaternions these conditions  can be rewritten  as
conditions for Fueter quaternion (anti) analyticity which, on
 the other hand,  coincide with the chiral parts of the Dirac massless equation in $E^4$.
It is shown in Ref.\cite{ds2} that massless spinors
can be quantized either as fermions or as bosons.
We use this result to  quantize anomalously the $U(1)$  ground state.
This is the key moment in the work --- we use fermions to describe instantons.
The change of the statistics `stabilizes' the instantons  allowing
non-trivial (anti) self-dual configurations to exist.
The idea that instantons change the statistics of some fields is not new ---
see, e.g.,  Refs.\cite{t4, t5, t6} for the $SU(2)$ case.
The topological reason for  the existence of these `anomalous' $U(1)$ instantons
is that using fermions we effectively change the
base of the $U(1)$ bundle from $E^4$ to its double covering space
which has non-trivial fundamental group.

Our second example is the pure $U(2)$ gauge model in $E^4$. We use the same strategy as in the $U(1)$ case. 
The aim  is to construct the largest possible vacuum  and to quantize this vacuum (anomalously).
The resulting theory describes the $U(1)\otimes SU(2)$ gauge interaction of the vacuum field and the obtained model looks very much like the lepton sector of the Standard Model. 

\section{Quantization around multiple vacuums}

In order to explain the specificity of the quantization procedure in the vicinity of non-unique vacuum we consider a (most general) model for a field $\vf$ with Lagrangean $L$.
 Let $\phi$ is a classical solution, i.e.,
\be
{\p L \over \p \vf }\vert_{\vf=\phi} = 0.\label{cls}
\ee
Let us expand the field $\vf$ in the vicinity of $\phi$
\be 
\vf = \phi +\eta. \label{env1}
\ee
Here $\eta$  is a (small) fluctuation around the classical solution usually called `quantum field'. 
Substituting eq.(\ref{env1}) into  the Lagrangean $L$ and keeping 
terms up to  second order  with respect to $\eta$ 
 we obtain $L \approx L'$ where
\be
L'(\phi,\eta)= L(\phi) + 
\hf\eta\left({\p^2 L\over \p\vf\p\vf}\vert_{\vf=\phi}\right)\eta.
\label{sol} 
\ee
The  operator $T =\hf(\p^2 L/ \p\vf\p\vf)\vert_{\vf=\phi}$ 
determines the propagator of the quantum field.
This is the `quantum evolution operator'.

If the vacuum is unique then eq.(\ref{env1}) is a simple change of variables
and the  path integral measures over $\vf$ and $\eta$ coincide.
However, an extra care is needed if the ground state is non-unique.
In this case the quantum evolution operator has zero modes.
Their existence is rather easy to demonstrate.
Suppose $\phi$ depends on some parameter, say $\alpha$.
Then the $\alpha$-derivative of the equation of motion (\ref{cls}) is
\be
0={\p\over\p\alpha}\left({\p L\over\p\vf}\vert_{\vf=\phi}\right) =
T {\p\phi\over\p\alpha}.
\ee
Therefore, $\p\phi/\p\alpha$ is a zero mode of the quantum evolution operator or in other words,  there is a gauge symmetry in the Lagrangean $L'$ with zero mode describing Goldstone boson.
The symmetry reflects the possibility  $\eta$ field to be
of the form $\p \phi / \p\alpha$ which corresponds to a vacuum to vacuum transformation and has to be excluded by proper redefinition of the path integral measure for the model with the Lagrangean (\ref{sol}).
The models with the Lagrangean $L(\vf)$ and $L'(\phi,\eta)$ are equivalent, and therefore
\be 
\int D\vf \;e^{-\int L}= \int\tilde{D}\eta\;\tilde{D}\phi \; e^{-\int L'}. \label{env2}
\ee
Eq.(\ref{env2}) just says that the transformation $\{\vf\}\rightarrow\{\phi, \eta\}$ is a kind of change of variables.
Here $\tilde{D}\eta$ is the standard measure in a gauge theory.
It incorporates the gauge fixing term $\chi$ (needed  to ensure that the quantum fluctuations are independent of the vacuum ones) and the corresponding  Faddeev-Popov determinant $\Delta$
\be 
\tilde{D}\eta= \delta(\chi)\Delta D\eta
\ee

The integration over the measure  $\tilde{D}\phi$ effectively sums different vacuums.
For instance, when $\alpha$ is a single  parameter
\be 
\tilde{D}\phi =  d \alpha/n(\alpha)\label{me3}
\ee
where $n(\alpha)$ is the norm of the zero mode.
Eq.(\ref{me3}) is the expression proposed in Refs.\cite{t1,t3} and then used
in the estimation of some non perturbative effects.
Eq.(\ref{me3}) is in agreement with the more general redefinition of the functional integral \cite{bd} as a sum over eigenvalues.

When the vacuum is not known explicitly but is  defined implicitly as a solution of some differential operator $\fd$ (which is {\it not} $\delta L / \delta\vf $) 
\be 
\fd \phi = 0 \label{vae}
\ee
then, in order to take into account all possible vacuums, the path integral measure $\tilde{D}\phi$ is
\be 
\tilde{D}\phi = \delta(\fd\phi) D\phi. \label{env3}
\ee
Using Lagrange multipliers  the delta function in eq.(\ref{env3}) can be 
represented as a part of the Lagrangean.
Finally we obtain the following Lagrangean for our model in the vicinity of the vacuum $\phi$ which satisfies the eq.(\ref{vae}): 
\be
L''=\bar\phi\fd\phi+L(\phi,\eta) + \mathrm{gauge \; fixing} + \mathrm{ghosts}.
\ee
Now, $\eta$, $\phi$ and $\bar\phi$ are independent fields.

We want to stress that it does not follow automatically from the above considerations that when the Lagrangean $L$ possesses {\it ad initium} gauge symmetry then $L'$ has bigger gauge symmetry. 
The Lagrangean (\ref{sol}) can  have the same symmetry as $L$ if $\p\phi/\p\alpha$ coincides with some of the zero modes due to the initial gauge freedom.
The gauge symmetry of $L'$ could be even smaller  than the symmetry of $L$
if $\phi$ is gauge non-invariant.
In any case one has to determine exactly the symmetry of the quantum evolution operator and to use adequate gauge conditions.

\section{Quaternions and Quaternion analyticity}

Hereafter we shall use intensively quaternions in our calculations.
In this section we just fix our notations.
Let us remind the definition of quaternion number $\qc \in \mathbbm{H}$
\be
\qc = h_\mu e_\mu \label{qn}
\ee
where $h_\mu,\;\;\; \mu= 0,..,3$ are four real numbers and
$e_\mu$ are four non-commutative quaternion units such that:
\bea
e_0 e_0 &=& e_0 \nn \\
e_i e_0 &=& e_0 e_i = e_i  \;\;\; i=1,2,3\nn \\
e_i e_j &=& - \delta_{ij} e_0 + \epsilon_{ijk} e_k. \label{qdef}
\eea
Here $\epsilon_{ijk}$ is the totally anti symmetric third rank tensor.
As a consequence of eqs.(\ref{qdef}) we can aways set the quaternion unit $e_0$ to $1$, so that $\qc = h_0 + h_i e_i$. 
There is a natural operation of conjugation in $\mathbbm{H}$.
We  denote by $\bqc$ the quaternion  conjugated to $\qc$ 
\be
\bqc = h_0 - h_i e_i.\label{cqn}
\ee

The so called Fueter analyticity is the quaternion analog to the complex
analyticity.
To define it we need two first order differential operators $\dc$ and $\bdc$
\bea
\dc&=& \p_0 + e_i \p_i \label{fd}\\
\bdc&=&\p_0 - e_i \p_i.\label{afd}
\eea
The function $\fc$ is called Fueter analytic if it satisfies 
the following equation
\be
\dc\fc =0.
\ee
If the function  satisfies the equation
\be
\bdc\fc =0
\ee
then the function is Fueter anti analytic.

Mainly we use the following representation of the quaternion units 
\be
e_0  =  \ou,\;\;
e_k  =  -i\sigma_k, \;\;\; k=1,2,3. \label{e-rep}
\ee
In eqs.(\ref{e-rep}) $\ou$ is the $2\times 2$ identity matrix and $\sigma_k$ are the Pauli matrices.
In a moment we shall need a  representation of $\mathbbm{H}\otimes\mathbbm{H}$ as well.
Two sets of commuting quaternion units $\{ e_\mu \}$ and $\{\xi_\mu\}$ are used in this case. The units $e_\mu$ are realized as in eqs.(\ref{e-rep}) and for $\xi_\mu$ we use another (real) representation with $4\times 4$ ($2\times 2$ block) matrices 
\be
\xi^0 = \left(
\begin{array}{cc} 
\ou & 0\\ 0 & \ou
\end{array}\right),  
\xi^1 = \left(
\begin{array}{cc} 
0& -\iu\\ -\iu& 0
\end{array}\right),
\xi^2 = \left(
\begin{array}{cc} 
0& -\ou\\  \ou & 0
\end{array}\right),
\xi^3 = \left(
\begin{array}{cc} 
-\iu & 0\\ 0 & \iu
\end{array}\right). \label{x-rep}
\ee
Here $\iu=-i\sigma_2$.
Together eqs.(\ref{e-rep},\ref{x-rep}) guarantee that we have a
well defined representation of $\mathbb{H}^2$.

\section{The pure $U(1)$ gauge model}

The quantization of the $U(1)$ (Electromagnetic) vacuum \cite{m1} is in the very heart of our discussion and we shall try to explain it in all possible details.
We consider the pure Electrodynamics  in four-di\-men\-sion\-al Euclidean space-time $E^4$.
The model is defined by the following action
\be
{\mathbf A} = - \int \of F_{\mu\nu}F_{\mu\nu}. \label{pure}
\ee
Here $F_{\mu\nu}$ is the field strength tensor for the
electromagnetic potential $A_\mu$;
 $F_{\mu\nu}=\partial_\nu A_\mu - \partial_\mu A_\nu$.
Having the electromagnetic potential  $A_\mu$ 
we can  associate to it two (conjugated) quaternion functions $\ac$ and $\bac$
\bea
 \ac &=&  A_0 + A_i e_i \label{qemf}\\
\bac &=& A_0 - A_i e_i. \label{cqemf}
\eea
The Fueter operators (\ref{fd},\ref{afd}) on the functions $\ac$ and $\bac$ give the following expressions:
\bea
\dc\bac &=& \partial_\mu A_\mu +
(-\partial_0 A_i + \partial_i A_0 - \epsilon_{ijk}\partial_j A_k) e_i \\
\bdc\ac &=& \partial_\mu A_\mu +
(\partial_0 A_i - \partial_i A_0 - \epsilon_{ijk}\partial_j A_k) e_i
\eea
Therefore, if the electromagnetic potential is Fueter anti analytic, i.e., if
\be
\bdc\ac =0, \label{sdu}
\ee
then we have self-dual configuration which in addition satisfies the Lorentz gauge condition.
On the other hand if the potential is Fueter analytic
\be
\dc\bac =0 \label{asdu}
\ee
then the field is anti self-dual plus again the Lorentz condition.

It is known that in the pure gauge theories the (anti) self dual field configurations are solutions of the equations of motion.
So, the idea is to use operators $\dc$ and $\bdc$ at the place of the operator $\fd$ in eq.(\ref{vae}).
We know that there are not nontrivial (anti) self dual $U(1)$ fields in $E^4$ but we just want to see what happens when we fairly apply the procedure outlined in section 2.
So, we formally consider a fixed  self-dual electromagnetic potential $A^+$ and an anti self-dual one $A^-$.
Both $A^+$ and $A^-$ are solutions of the classical equations of motion and  (because Maxwell's equations are linear) so is their sum.
Thus the most general vacuum we can construct is $\phi= A^+ + A^-$ and we expand the Electromagnetic potential around it
\be
A_\mu = A^+_\mu + A^-_\mu + A'_\mu. \label{dec}
\ee
Let us denote by $F^+$, $F^-$ and $F'$  the field strengths 
which correspond to the potentials $A^+$, $A^-$ and $A'$ respectively.
The following well known relations are fulfilled for these quantities 
\bea
F^+_{\mu\nu}F^-_{\mu\nu}&=&0\nn\\
F^+_{\mu\nu}F^+_{\mu\nu}&=& 2 \partial_\lambda \left(
\epsilon_{\lambda\mu\nu\rho} A^+_\mu \partial_\nu A^+_\rho\right) \nn\\
F^-_{\mu\nu}F^-_{\mu\nu}&=& - 2 \partial_\lambda \left(
\epsilon_{\lambda\mu\nu\rho} A^-_\mu \partial_\nu A^-_\rho\right).\label{asre}
\eea
It is easy to show using eqs.(\ref{asre}) that the electromagnetic action (\ref{pure})
in the vicinity of the ground state $A^++A^-$  takes the form
\be
{\mathbf A}=- \int \of F'_{\mu\nu}F'_{\mu\nu}. \label{aacs}
\ee
As a consequence the quantum evolution operator
 coincides with the Maxwell operator. 
There are no extra zero modes and the gauge symmetry group is $U(1)$.
Certainly, in order to obtain a well-defined transition amplitude,
we have to choose a gauge fixing function.
Once again, because of eq.(\ref{aacs}) the gauge fixing function can be each 
of the standard  ones used in Electrodynamics
(now written for the field $A'$).
Therefore we can view $A'$ as the standard electromagnetic potential
 and eq.(\ref{aacs}) as the standard pure electromagnetic action.

Another consequence of eq.(\ref{aacs}) is that 
because the action $\mathbf A$ does not depend on $A^+$ and $A^-$ the
contribution  of the vacuum modes to the transition amplitude will be
a multiplicative constant which gives the number of
 different (anti) self-dual configurations.
In order to find this number we have to clarify 
what `different (anti) self-dual configurations' means,
which is the question for the gauge freedom in $A^+$ and $A^-$.
Writing eq.(\ref{dec}) we have assumed that $A'$ is a connection as the field $A$. 
The difference of two connections is a tensor (invariant in our $U(1)$ case), and so the proper handling of the expansion (\ref{dec}) requires the vacuum to be gauge invariant, i.e.
\be
\delta_\epsilon (A^+_\mu + A^-_\mu)=0.\label{gv}
\ee
Here $\delta_\epsilon $ denotes the usual $U(1)$ gauge variation with 
parameter $\epsilon$.
However, even if $A^+ + A^-$ satisfies eq.(\ref{gv})
there is still room for a new gauge freedom.
In general, if a field $B_\mu$ is represented as a sum of two independent fields
$B_\mu=B^+_\mu + B^-_\mu$, then a very large Stuckelberg symmetry emerges
$\delta B^\pm_\mu=\pm C_\mu$, 
where $C_\mu$ is an arbitrary vector field.
In our case, because of duality properties of $A^+$ and $A^-$, only the following remnant  of the Stuckelberg symmetry is allowed
\be
\delta'_\zeta (A^\pm_\mu) = \pm \partial_\mu\zeta \label{agi}
\ee
($\zeta$ is an arbitrary function).
Thus the decomposition (\ref{dec}) introduces a new gauge freedom in the model
which is independent from the initial one.
But we do not want it, so we have to fix it ensuring that
\be
\delta'_\zeta (A^+_\mu - A^-_\mu)=0.\label{sgv}
\ee
So, we have to impose two conditions (\ref{gv}) and (\ref{sgv}) 
on two independent combinations of the potentials $A^+$ and $A^-$
which is equivalent to fixing the gauge separately in both the
self-dual and anti self-dual potentials.
We choose as gauge conditions 
$$\partial_\mu A^\pm_\mu =0.$$

Up to now the result of our a little long considerations is that 
 the requirements for Fueter anti-analyticity for $\ac^+$ and 
Fueter analyticity for $\bac^-$   specify the possible $U(1)$ vacuums.
According to eq.(\ref{vae})
 the integration measure over self-dual configurations  must be something like 
\be
\delta(\bdc\ac^+)\;\; D A^+ \label{measure}
\ee
with an analogous expression for the measure over the anti self-dual potentials. 

Here is the moment to use a concrete representation of the quaternion units\footnote{
The use of representation (\ref{e-rep}) can be viewed as a change of variables such that the vector is described by bispinor.}.
The explicit form of the quaternion $\qc$ in representation (\ref{e-rep}) is
\be
\qc  = \left(
\begin{array}{cc} 
a & - \bar b\\b & \bar a
\end{array}\right) \label{rpr}
\ee
where $a = h_0 - i h_3$, $b=h_2 - i h_1$ and $\bar a$, $\bar b$ 
denoting the complex conjugates of $a$ and $ b$.

An important consequence of eq.(\ref{rpr}) is that 
if for some normalized Weyl spinor $v$, 
say $v=\left( \begin{array}{c} 1\\0\end{array}\right)$,
we know $\qc\cdot v $, then we know $\qc$ itself.
This fact has to be taken into account when we write
the integration measure.
Using the representation (\ref{e-rep}) we obtain   
in eq.(\ref{measure}) a delta function of  a matrix which has to 
be understood as a product of delta functions of each
matrix entry.
This gives four complex conditions for $A^+$ which exceeds the 
correct number.
In order to get two complex conditions for $A^+$ 
 we have to pick up a constant spinor $v$ 
which to multiply $\dc\ac^+$ on the right and to use  $\dc\ac^+\cdot v$ 
as an argument of the delta function in eq.(\ref{measure}).
Let us define the spinors $\psi^+$ and $\psi^-$ as follows
\bea
\psi^+ &=& \ac^+\cdot v\nn\\
\psi^- &=& \bac^-\cdot u .\label{ano}
\eea
Then the transition amplitude for the  pure Electrodynamics
 plus vacuum contributions and  without gauge freedom of the quantum field
 to be fixed takes the form
\be
S=\int D \psi^+ D \psi^- D A' \;
\delta(\bdc\psi^+)\delta(\dc\psi^-)
\exp\left\{- \of\int F'_{\mu\nu}F'_{\mu\nu}\right\}.\label{ta1}
\ee
The two delta functions in eq.(\ref{ta1}) can be represented
as one.
In order to do this 
we introduce a four-dimensional Dirac spinor $\psi$ 
as  direct sum of the two Weil spinors $\psi^+$ and $\psi^-$ 
\be
\psi=\left(\begin{array}{c} 
\psi^-\\\psi^+\end{array}\right)\label{dspin}
\ee
We also use the diagonal $\gamma^5$ representation of the four-dimensional
 Euclidian gamma matrices
\be
\gamma_0=\left(\begin{array}{cc} 
0 & 1\\1 & 0\end{array}\right),\;\;\;
\gamma_i=\left(\begin{array}{cc} 
0 & i\sigma_i\\-i\sigma_i & 0\end{array}\right),\;\;\;
\gamma_5=\left(\begin{array}{cc} 
-1 & 0\\0 & 1\end{array}\right).
\ee
In this representation the  Dirac operator has the form
\be
\partial_\mu \gamma_\mu = 
\left(\begin{array}{cc} 
0 & \partial_0 + i \partial_k \sigma_k\\
\partial_0 - i \partial_k \sigma_k& 0\end{array}\right) = 
\left(\begin{array}{cc} 
0 & \bdc\\ \dc & 0\end{array}\right). \label{dslash}
\ee
The operators $\bdc$ and $\dc$ which appear in eq.(\ref{dslash})
are those defined by eqs.(\ref{fd}, \ref{afd}) taken  in the
representation (\ref{e-rep}).
Using eqs.(\ref{dspin}, \ref{dslash}) we get:
\be
\delta(\bdc\psi^+)\delta(\dc\psi^-)=\delta(\partial_\mu \gamma_\mu \psi).
\label{fdf}
\ee
We use  Lagrange multipliers to present the delta function (\ref{fdf}) as part of the action.
Because  of the hermiticity of the action, the Lagrange multipliers
have to form a spinor, conjugated to the spinor $\psi$ \cite{ds2}.
Thus we get the following expression for the transition amplitude:
\be
S=\int D\bar\psi D\psi D A' 
\exp\left\{- \int i \bar\psi \partial_\mu \gamma_\mu \psi +
\of F'_{\mu\nu}F'_{\mu\nu} \right\}.\label{ta2}
\ee
Here is the moment to apply further the results of Ref.\cite{ds2} and to quantize anomalously (as fermions) the field $\psi$.
This is the most important step in our work. 
By using fermions to describe (anti) self dual $U(1)$ field configurations we actually take into account new field modes.
This was not our initial intention but the form of eq.(\ref{ta2})  suggests  very strongly the use of Fermi--Dirac statistics.
The anomalous quantization  effectively changes the trivial $U(1)$ bundle over $E^4$ to a one whose base $B$ is a double cover of $E^4$ and thus is with nontrivial fundamental group.

There is a little flow in our recipe to pass from quaternions  to spinors.
Unfortunately, following this  recipe we face a problem 
which is due to the gauge non-invariance of constant spinor $v$ we use: the quantity $\dc\ac^+\cdot v$ is neither gauge invariant, nor gauge covariant.
So, the price we have to pay in order to have the correct number of constraints
is gauge non-invariance of the obtained expression.
As a result the action in eq.(\ref{ta2}) is  gauge non-invariant as well.
However, we have started with gauge invariant action, and we have to restore
this invariance in eq.(\ref{ta2}).
We do this in the standard way, namely prolonging the derivatives with $A'$
thus obtaining
\be
S=\int D \bar\psi D \psi D A' 
\exp\left\{- \int \bar\psi( i \partial_\mu + A'_\mu )\gamma_\mu \psi +
 \of F'_{\mu\nu}F'_{\mu\nu} \right\}.\label{ta3}
\ee
It is possible to get eq.(\ref{ta3}) directly without
passing through eq.(\ref{ta2}).
For this purpose we have to use in eq.(\ref{ano}) a gauge invariant spinor $w$
instead of the constant but gauge non-invariant spinor $v$.
Let us define $w$ as follows:
\be
w = e^{-i \phi\lb\Gamma\rb} v.\label{gii}
\ee
Here $\phi\lb\Gamma\rb$ is a phase which compensates the gauge
transformation of $v$.
The phase $\phi\lb\Gamma\rb$ is non-integrable \cite{dirac}, i.e. 
it is not a function with definite value in each space-time point $x$, 
but instead it is a multi-valued functional depending on the path $\Gamma$ along which
we reach the point $x$. 
However, the phase $\phi\lb\Gamma\rb$ possesses well defined derivatives
\be
\partial_\mu \phi\lb\Gamma\rb = A'_\mu .\nn
\ee
Formally, $\phi\lb\Gamma\rb$ can be written as an integral over 
path $\Gamma$ to the point $x$
\be
\phi = \int_\Gamma d l. A'.
\ee
When we use $w$ to reduce the number of 
constraints on $A^+$ we get
\bea
\bdc\ac^+\cdot w & = &
\bdc (e^{-i \phi\lb\Gamma\rb} \ac^+\cdot v) \nn\\
&=& e^{-i \phi\lb\Gamma\rb} (\bdc - i \ac' )\psi^+ \label{gis}
\eea 
and this gauge invariant expression we have to put into  the delta function
which defines the measure over self-dual field configurations.
Representing the delta function as an exponent, the phase 
$ e^{-i \phi\lb\Gamma\rb}$ 
in eq.(\ref{gis}) will be compensated by the (conjugated to it) 
phase of the Lagrange multipliers.
Thus the phase $\phi\lb\Gamma\rb$ does not appear at all in the Lagrangean.
Applying the same procedure to the measure which counts anti self-dual
fields, we get for the transition amplitude directly
the gauge invariant expression (\ref{ta3}).

\section{The pure $U(2)$ gauge model}

The action for the pure $U(2)$ gauge  model in $E^4$ is
\be
{\mathbf A} =-\of \int  F_{\mu\nu}^\alpha F_{\mu\nu}^\alpha.
\label{sa}
\ee
Here $\alpha=0,1,2,3$ are the $U(2)$ indexes 
($\alpha=0$ is the $U(1)$ index and $\alpha=1,2,3\;$ are the $SU(2)$ ones
in the decomposition $U(2)=U(1)\times SU(2)$).
In our notations the $U(1)$ charge is $1$ and the $SU(2)$ charge is $g$,
so that the corresponding field strengths are 
\bea
F_{\mu\nu}^0 &=& \partial_\mu A_\nu^0 - \partial_\nu A_\mu^0.\label{u1fs}\\
F_{\mu\nu}^a &=& \partial_\mu A_\nu^a - \partial_\nu A_\mu^a +
g \lb A_\mu, A_\nu\rb^a,\;\;a=1,2,3.   \label{su2fs}   
\eea

The idea again is to find a classical solution of the action (\ref{sa})
which to use as a vacuum.  
Again  we use (anti) self-dual fields to build the vacuum.
We shall need two types of mutually commuting ($e$- and $\xi$-) quaternions. 
We connect the $e$-quaternions to the space-time while the
$\xi$-quaternions are connected to the internal $U(2)$ space.
Thus, having a potential  $A_\mu^\alpha$ we  construct out of it 
four $e-$quaternion functions 
$
\ac^\alpha,
$
 four $\xi-$quaternion functions 
$
\af_\mu
$
and one bi-quaternion function
$
 \ab :
$
\bea
\ac^\alpha &=& e_\mu A_\mu^\alpha\\
\af_\mu    &=& \xi^\alpha A_\mu^\alpha\label{crivoa}\\
\ab        &=& \xi^\alpha e_\mu A_\mu^\alpha.
\eea
We shall use also the $e-$conjugated to $\ac$ and $\ab$  functions
which we denote $\bac^\alpha$ and $\bab$ respectively
\bea 
\bac^\alpha & = & \bar{e}_\mu A_\mu^\alpha\nn\\
\bab & = & \xi^\alpha \bar e_\mu A_\mu^\alpha.\nn
\eea
Note, that there is no $\xi-$quaternion conjugation in the
definition of $\bab$.

After some algebra, we get that the equation  
\be 
(\bdc + \ghf \bab)\ab = 0  \label{sdu2}
\ee
describes a self-dual field configuration in the following gauge
\bea 
\p_\mu A_\mu^0 + \ghf \left((A^0)^2-(A^a)^2\right)&=&0\nn\\
\p_\mu A_\mu^a + g A^0\cdot A^a &=& 0. \label{gg}
\eea
The equation
\be 
(\dc + \ghf \ab)\bab = 0  \label{asdu2}
\ee
describes an anti self-dual field  also in the gauge (\ref{gg}).
The operators $\dc$ and $\bdc$ in eqs.(\ref{sdu2},\ref{asdu2}) 
should be understood as those defined by eqs.(\ref{fd},\ref{afd})
multiplied by $\xi^0$ 
(as it is usual for the definition of the covariant derivative).
We want to emphasize that the gauge (\ref{gg}) is not unique or preferred --- 
any gauge condition can be obtained adding an arbitrary 
$\xi-$quaternion function to eqs.(\ref{sdu2},\ref{asdu2}).
We shall use this freedom below.

Using Fueter anti analytic functions we can
construct a rather general solution of eqs.(\ref{sdu2}).
Let $\hat{\ab}$ be a $\xi$-valued Fueter anti analytic function
(four solutions of the eq.(\ref{afd}) combined into a single bi-quaternion).
Consider the function
\be 
\ab(x) = e^{-\ghf\int_\Gamma \hat{\af}\cdot dl} \hat{\ab}(x) \label{gso}
\ee
where $\Gamma$ is some path to the point $x$ and the notation $\af$
is introduced in eq.(\ref{crivoa}).
It is easy to show that $\ab$ thus defined is a solution of eq.(\ref{sdu2}).
An analogous solution of the eq.(\ref{asdu2}) can be constructed out of four Fueter analytic functions.

Eq.(\ref{gso}) gives a rather formal solution which is hard to use explicitly.
However it encodes a very important information, namely
 that four Fueter anti analytic functions parameterize a $U(2)$ self dual field\footnote{Here we shall not discuss the question whether eq.(\ref{gso}) gives one-to-one correspondence between $U(2)$ instantons and Fueter anti analytic function.
For a moment we simply assume that this is a parameterization of (some of) the solutions of eq.(\ref{asdu2}).}.
We know from the previous section how to deal with Fueter  analytic and anti analytic functions and what is their particle interpretation.
All we have to do now is to find the maximal classical solution of the problem.
But the things are not so simple because there are two very important differences between the Abelian and non Abelian gauge models.
First, 
contrary to the $U(1)$ case, the Yang--Mills equations are non liner and so, in general, the sum of two solutions is not a solution.
An obvious exception of this rule is if we add a solution which belongs to the  center of the gauge algebra.
In our case this center is $u(1)$.
As a consequence we can add to the $U(2)$ instanton (\ref{gso}) only an $U(1)$ anti instanton (\ref{asdu}) or to add to the  $U(2)$ anti instanton an $U(1)$ instanton. 
Second, the Yang--Mills vacuum is not gauge invariant as the Electromagnetic one, it is gauge covariant.
So, our arguments concerning eqs.(\ref{gv}--\ref{sgv}) have to be changed.
Now it is absolutely necessary the vacuum to be a difference of two connections in order to transform correctly.
But the $SU(2)$ part of the eq.(\ref{gso}) can not be superposed with other instanton or anti instanton solution.
Therefore we can not use eq.(\ref{gso}) as a part of the vacuum.  

In order to illuminate the situation we consider a simpler then eq.(\ref{gso}) solution of the self duality equations.
A suitable candidate  is a field for which the
nonlinear term in eq.(\ref{sdu2}) (or eq.(\ref{asdu2})) vanishes.
This effectively reduces the non-Abelian (anti) self-dual conditions to Abelian ones which, in their turn, are the Fueter (anti) analyticity conditions.
Two such solutions are easy to find\footnote{
A change of the gauge condition (\ref{gg}) is required for both solutions.}.
{\bf Solution (a): }
Let $\ac^0\neq 0$ and $\ac^a=0$ for $a=1, 2, 3$. 
In addition we require that either  $\ac^0$  satisfies  eq.(\ref{sdu}) or $\bac^0$ satisfies eq.(\ref{asdu}).
This is the $U(1)$ solution we have discussed in the previous section.
{\bf Solution (b): } Fix a $su(2)$ index, say $a$ and let $\ac^\alpha=0\;$ $\forall \alpha\neq a$ while $\ac^a$ is a non zero solution of eq.(\ref{sdu}), or $\bac^a$ is a  solution of eq.(\ref{asdu}).
This solution, as the previous one, describes an $U(1)$ vacuum.
However, there are two important differences between solutions (a) and (b).
First, they are associated with two different $u(1)$ algebras --- 
solution (a) is related to the proper $u(1)$ subalgebra of $u(2)$,
while  solution (b) is  related to the Cartan subalgebra of $su(2)$.
Therefore, these solutions have different charges. 
Second, there is a global $SO(3)$ covariance for solution (b), which resembles the situation with solution (\ref{gso}). 
If $\ac^a$ is a solution, so is the field 
\be
\ac '^a = U^{ab}\ac^b\label{so3i}
\ee
where $U^{ab}$ is a $SO(3)$ matrix.

Solutions (a) and (b) can be freely combined with each other
and their superposition still will be a solution of the equation of motion.
(We want to recall that the situation was different when we use eq.(\ref{gso}) to construct the vacuum.)
Therefore, the largest $U(2)$ vacuum, we can construct out of solutions (a) and (b)
describes  two left and two right massless spinor fields.
There is no left--right asymmetry in this case.
We can speculate a little with the $SO(3)$ symmetry (\ref{so3i}).
It indicates that this vacuum is not unique and the space of the vacuums can be span by three basic vectors. 
We can numerate them simply by $v^1, v^2, v^3$ or, if you prefer, $e, \mu, \tau$.
Note that only one of these vacuums can be used as an  asymptotic state, or in other words there is only one stable particle.

\section{Conclusions}

It is known that the difference between two connections transforms as a matter field under the action of the gauge group.
What we argue here is that the matter {\it is} a difference of two gauge connections. 

We show that the (anomalous) quantization of the non trivial vacuums of the pure Electromagnetic model leads to full fledged QED in which the spinor field is not put `by hands' but emerges `naturally'.
We apply the same approach to the  pure $U(2)$ Yang--Mills theory.
It is tempting to associate the constructed vacuums with the  lepton sector of the Standard model.
However a lot of additional work is needed in order to do this.
It has to be clarified the physical model behind the vacuum based on eq.(\ref{gso}). 
This solution seems the most general one but the vacuum constructed from simpler field configurations are qualitatively different from those based on  eq.(\ref{gso}).
Another important problem we have to solve is to find a selection rule for the vacuum charges.
If the matter is a non trivial gauge vacuum its interaction constants can not be arbitrary.
However, at the  moment the vacuum charges are not determined.
We have to find some additional explanation of these features
if we want to use the anomalous quantized vacuum as the origin of the leptons.

\section*{Acknowledgement}
The work is suppurted by BNSF grant 2-288.

\end {document}